\documentclass[11pt,a4paper]{article}
\usepackage{amsfonts}

\def\be{\begin{equation}}
\def\ee{\end{equation}}
\def\bea{\begin{eqnarray}}
\def\eea{\end{eqnarray}}
\def\onehalf{{\textstyle{\frac{1}{2}}}}

\def\rt{{\mbox{\tiny{(T)}}}}
\def\rk{{\mbox{\tiny{(K)}}}}

\def\a{{\mbox{\footnotesize{a}}}}
\def\s{{\mbox{\footnotesize{s}}}}
\begin{document}
\noindent
{\Large \bf de Sitter Relativity: a New Road to Quantum Gravity?}
\vskip 0.7cm
\noindent
{\bf R. Aldrovandi and J. G. Pereira}\\ 
{\it Instituto de F\'{\i}sica Te\'orica, Universidade Estadual Paulista \\
Rua Pamplona 145}, {\it 01405-900 S\~ao Paulo, Brazil}

\vskip 0.8cm 
\noindent
{\bf Abstract~} {\small The Poincar\'e group generalizes the Galilei group for {\it high--velocity} kinematics. The de Sitter group is assumed to go one step further, generalizing Poincar\'e as the group governing {\it high--energy} kinematics. In other words, ordinary special relativity is here replaced by de Sitter relativity. In this theory, the cosmological constant  $\Lambda$ is no longer a free parameter, and can be determined in terms of other quantities. When applied to the whole universe, it is able to predict the  value of $\Lambda$ and to explain the cosmic coincidence. When applied to the propagation of ultra--high energy photons, it gives a good estimate of the time delay observed in extragalactic gamma--ray flares. It can, for this reason, be considered a new paradigm to approach the quantum gravity problem.}




\section{Introduction}

Low energy physics is governed by Newtonian mechanics, whose underlying kinematics is ruled by the Galilei group. For higher energies, which involve higher velocities, Galilei relativity fails and must be replaced by Einstein special relativity, whose underlying kinematics is ruled by the Poincar\'e group. From the kinematic point of view, Poincar\'e relativity can be viewed as describing the implications to Galilei relativity of introducing a fundamental velocity scale --- the speed of light $c$ --- into the Galilei group. Conversely, Galilei relativity can be obtained from Poincar\'e's by taking the formal limit of the velocity scale going to infinity (non-relativistic limit).

Now, there are theoretical and experimental evidences that, at ultra--high energies, also Poincar\'e relativity  fails to be true. The theoretical indications are related to the physics at the Planck scale, where a fundamental length parameter --- the Planck length $l_P$ --- naturally shows up. Since a length contracts under a Lorentz boost, the Lorentz symmetry is usually supposed to be broken at this scale~\cite{lorentzX}. The experimental evidences come basically from the propagation of very--high energy photons, which seems to violate ordinary special relativity. More precisely, very--high energy extragalactic gamma--ray flares seem to travel slower than lower energy ones~\cite{magic}. If this comes to be confirmed, it will constitute a clear violation of special relativity.

The above indications suggest that we should look for another special relativity, which would rule the kinematics at ultra--high energies.\footnote{Several attempts have been made to construct such a theory. The relevant literature can be traced back from the papers cited in Ref.~\cite{dsr}.} From the point of view of the algebraic hierarchy described above, the most natural generalization towards ultra--high energy kinematics would be to replace Poincar\'e special relativity by a de Sitter relativity~\cite{dssr,guoatall}. Algebraically, this theory describes the implications to Galilei relativity of introducing {\em both} a velocity and a length scales in the Galilei group. In the formal limit of the length-scale going to infinity, the de Sitter group contracts~\cite{inonu,gil} to the Poincar\'e group~\cite{gursey}, in which only the velocity scale is present. A further limit of the velocity scale going to infinity leads Poincar\'e to Galilei relativity. It is interesting to observe that the order of the group expansions (or contractions) is not important. If we introduce in the Galilei group a fundamental length parameter, we end up with the Newton-Hooke group~\cite{nh}, which describes a Galilean relativity in the presence of a cosmological constant. Adding to this group a fundamental velocity scale, we end up again with the de Sitter group, whose underlying relativity involves both a velocity and a length scales. Conversely, the low--velocity limit of the de Sitter group yields the Newton-Hooke group, which contracts to the Galilei group in the limit of a vanishing cosmological constant.

Considering that the de Sitter group naturally incorporates an invariant length--pa\-rame\-ter --- in addition to the speed of light --- de Sitter relativity can be interpreted as a new example of the so called {\em doubly special relativity}~\cite{dsr}. There is a fundamental difference, though: whereas in all doubly special relativity models the Lorentz symmetry is violated, in de Sitter relativity it remains as a physical symmetry~\cite{lorentz}. It is important to mention that one drawback of the usual doubly special relativity models is that they are valid only at the energy scales where ordinary special relativity is supposed to break down,\footnote{This restriction is known as the ``soccer-ball problem'' \cite{soccer}.} giving rise to a kind of patchwork relativity. On the other hand, de Sitter relativity is found to be invariant under a simultaneous re-scaling of mass, energy and momentum~\cite{dssr}, and is consequently valid at all energy scales~---~it is a ``universal'' relativity. This is a very important property shared by all fundamental theories.

Taking into account the above considerations, the purpose of this paper is to explore the consequences of replacing ordinary special relativity by de Sitter special relativity. We begin by presenting, in section 2, the fundamentals of the de Sitter special relativity. If special relativity changes, general relativity must change accordingly. The consistency of these modifications is examined in section 3. As illustrations of possible applications of de Sitter special relativity, we use it in section 4 to re--analyze the cosmological constant problem, and in section 5 we study the implications of the theory for the propagation of very--high energy photons. Finally, in section 6, we present some concluding remarks.

\section{Fundamentals of de Sitter Relativity}
\label{sec:dsr}

According to de Sitter relativity, the local symmetry of spacetime is not ruled by Poincar\'e, but by the de Sitter group. Now, in order to exhibit de Sitter symmetry, a physical phenomenon must modify the local structure of spacetime in such a way that {\it the region in which it takes place becomes a de Sitter spacetime.}\footnote{This hypothesis has already been considered by F.\ Mansouri in a different context~\cite{mansouri}.} To comply with this requirement, in addition to the usual gravitational field, any physical system must also engender a local de Sitter field, whose intensity --- described by the local value of the ``cosmological'' term $\Lambda$, ultimately an energy density --- is proportional to the energy density of the physical system. The natural question then arises: how can ordinary matter give rise to a cosmological term? The answer to this question is not simple, and requires a thorough analysis.

\subsection{Conformal Transformations and the Cosmological Term}

In one of its versions, the strong equivalence principle states that, in the presence of a gravitational field, it is always possible to find a local coordinate system in which the laws of physics reduce to those of special relativity. In this local coordinate system, therefore, the kinematics is governed by the Poincar\'e group. This version of the equivalence principle, therefore, is consistent with ordinary special relativity, whose underlying spacetime is the Minkowski space
\be
M = {\mathcal P}/{\mathcal L},
\ee
the quotient between Poincar\'e and Lorentz groups. It is a solution of the sourceless Einstein equation\footnote{We use the Greek alphabet $(\mu, \nu, \rho, \dots = 0,1,2,3)$ to denote indices related to spacetime, and the first half of the Latin alphabet $(a,b,c, \dots = 0,1,2,3)$ to denote algebraic indices, which are raised and lowered with the Minkowski metric $\eta_{ab} = \mbox{diag}~(+1,-1,-1,-1)$. The second half of the Latin alphabet $(i,j,k, \dots = 1,2,3)$ is reserved for space indices.}
\be
R_{\mu \nu} - \onehalf \, g_{\mu \nu} \, R = 0.
\ee
An important property of the Minkowski spacetime is that it is transitive under spacetime translations. The invariance of a physical system under spacetime translations leads to the conservation of its corresponding Noether current, the energy-momentum tensor $T_{\mu \nu}$ which appears as the source in  Einstein's equation
\be
R_{\mu \nu} - \onehalf \, g_{\mu \nu} \, R = \frac{8 \pi G}{c^4} \, T_{\mu \nu}.
\label{ufee}
\ee

On the other hand, if the local kinematics is to be governed by the de Sitter relativity, the local symmetry group of spacetime changes from Poincar\'e to de Sitter, and consequently the strong equivalence principle must change accordingly. Its modified version states that, in the presence of a gravitational field, it is always possible to find a local coordinate system in which the laws of physics reduce to those of de Sitter special relativity, whose underlying spacetime is the de Sitter space
\be
dS(4,1) = {SO(4,1)}/{\mathcal L},
\ee
the quotient between de Sitter and Lorentz groups. Immersed in a five--dimensional pseudo-Euclidean space ${\bf E}^{4,1}$ with Cartesian coordinates $ (\chi^a, \chi^{4})$, it is defined by
\be
\eta_{a b} \, \chi^{a} \chi^{b} - (\chi^{4})^2 
= - \, l^2,
\label{dspace1}
\ee
with $l$ the so called de Sitter length--parameter (or ``pseudo--radius''). de Sitter space is a solution of the sourceless $\Lambda$-modified Einstein equation\footnote{According to our convention, the de Sitter spacetime ($\Lambda > 0$) has a negative scalar curvature ($R<0$). In this convention the scalar curvature has the same sign as the Gaussian curvature.}
\be
R_{\mu \nu} - \onehalf \, g_{\mu \nu} \, R - g_{\mu \nu} \, \Lambda = 0,
\label{usuee}
\ee
provided $\Lambda$ and $l$ are related by
\be
\Lambda = \frac{3}{l^2} \, .
\label{pureds}
\ee

Differently from Minkowski,  de Sitter spacetime is transitive under a combination of translations and proper conformal transformations~\cite{abap}. This property is easily seen in the coordinates $\{x^a\}$  obtained by  a stereographic projection from the de Sitter hyper-surface into a target Minkowski spacetime. The projection is defined by~\cite{gursey}
\be
\chi^{a} = \Omega(x) \, x^a \quad \mbox{and} \quad \chi^4 = -\, l \, \Omega(x) \left(1 + \frac{\sigma^2}{4 l^2} \right),
\label{stepro}
\ee 
where
\be
\Omega(x) = \frac{1}{1 - {\sigma^2}/{4 l^2}} \, , 
\label{n}
\ee
with $\sigma^2 = \eta_{ab} \, x^a x^b$ a Lorentz invariant quadratic interval. In such coordinates, the de Sitter metric has the form 
\be
{g}_{\mu \nu} = \Omega^2(x) \, \delta^a_{\mu} \, \delta^b_{\nu} \, \eta_{ab}  = \Omega^2(x) \, \eta_{\mu \nu},
\label{dsmetric}
\ee
showing clearly its conformally flat character. The generators of the de Sitter Lie algebra, on the other hand, are given by
\be
L_{ab} = \eta_{ac} \, x^c \, P_b - \eta_{bc} \, x^c \, P_a,
\ee
and
\be
L_{a4} = l P_a - (4l)^{-1} K_a,
\ee
where
\be
P_a = \partial_a \quad \mbox{and} \quad
K_a = \left(2 \eta_{ac} \, x^c x^b - \sigma^2 \, \delta_a^b \right) \partial_b
\ee
are, respectively, the generators of translations and proper conformal transformations. Generators $L_{ab}$ refer to the Lorentz subgroup, whereas the remaining $L_{a4}$ define the transitivity on the de Sitter spacetime.

To make contact with the Poincar\'e group, it is convenient to define~\cite{gursey}
\be
\pi_a \equiv \frac{L_{a4}}{l} = P_a - (4l)^{-2} K_a,
\ee
which are usually called de Sitter ``translation'' generators. From the algebraic point of view, therefore, the change from Poincar\'e to de Sitter is achieved by replacing ordinary translations $P_a$ by the de Sitter ``translations'' $\pi_a$. The relative importance of translation and proper conformal generators is determined by the value of $l$, that is, by the value of the cosmological term. We see in this way that a non--vanishing $\Lambda$ is directly related to the presence of conformal transformations in the spacetime transitivity generators --- or equivalently, in  the group governing the spacetime kinematics.

\subsection{The Local Cosmological Term}

We are now back to the question posed at the beginning of the section, which can now be rephrased in the form: given a physical system, how to obtain the associated cosmological term? To answer this question we remember first that, according to quantum mechanics, there is a lower limit for all physical quantities. For example, the smallest amount of an electromagnetic field, a photon, is determined by the Planck constant as a quantum of the field. In a similar fashion, the smallest possible length is the Planck length. Since in de Sitter relativity there is a free length parameter $l$, its minimum value will then be the Planck length $l_P = \sqrt{G \hbar/ c^3}$. Let us then consider a de Sitter spacetime with $l = l_P$, for which the corresponding cosmological term is
\be
\Lambda_P = \frac{3}{l^2_P}.
\label{L1}
\ee
Considering that a cosmological term represents ultimately an energy density, we define the Planck   energy density 
\be
\varepsilon_P = \frac{m_P \, c^2}{(4 \pi/3) l_P^3} \, ,
\ee
with $m_P = \sqrt{c \hbar/G}$ the Planck mass. In terms of $\varepsilon_P$, Eq.~(\ref{L1}) assumes the form 
\be
\Lambda_P = \frac{4 \pi G}{c^4} \, \varepsilon_P  \, .
\label{kineLambda0}
\ee
Now, the very definition of $\Lambda_P$ can be considered a particular, extremal case of a general expression relating the local ``cosmological'' term to the corresponding energy density of a physical system. Accordingly, to a physical system of energy density $\varepsilon$  will be associated the ``cosmological'' term 
\be
\Lambda = \frac{4 \pi G}{c^4} \, \varepsilon.
\label{kineLambda}
\ee
It is important to reinforce that the $\varepsilon$ appearing in this equation is not the dark energy density, but the matter energy density. For small values of $\varepsilon$, the local cosmological term $\Lambda$ will be small, spacetime will approach Minkowski, and de Sitter special relativity will approach ordinary special relativity, whose kinematics is governed by the Poincar\'e group.\footnote{We remark that the relation (\ref{kineLambda}) between the local value of $\Lambda$ and the energy density $\varepsilon$ is the same as that appearing in Einstein universe. See, for example, Ref.~\cite{narlikar}, page 104.}

\section{Consistency with General Relativity}

In order to comply with de Sitter relativity, any physical system must engender on space\-time a local cosmological term. This means that spacetime must present a local kinematic--related curvature. We have then to verify whether the presence of this kinematic curvature is consistent with general relativity, the theory that governs the spacetime dynamical curvature. 

\subsection{Conserved Source Currents}

Due to the transitivity properties of the de Sitter spacetime, de Sitter relativity naturally incorporates the conformal generators in the definition of spacetime transitivity. As a consequence, the conformal current will appear as part of the Noether conserved current, producing a change in the very notions of energy and momentum~\cite{vaxjo}. To see that, let us consider a general matter field with Lagrangian ${\mathcal L}_m$. Its action integral is
\be
S_m = \frac{1}{c} \int {\mathcal L}_m \, d^4x.
\ee
Under a local spacetime transformation $\delta x^\rho$, the change in $S$ is
\be
\delta S_m = -\,  \frac{1}{2 c} \int T^{\mu \nu} \, \delta g_{\mu \nu} \, \sqrt{-g} \, d^4x,
\label{Svar1}
\ee
where
\be
T^{\mu \nu} = -\,  \frac{2}{\sqrt{-g}} \, \frac{\delta {\mathcal L}_m}{\delta g_{\mu \nu}}
\label{syem}
\ee
is the symmetric energy--momentum tensor. Although the coefficient of the variation is the energy--momentum tensor, the conserved  quantity depends on the transformation $\delta x^\rho$. For example, invariance of the action under translations $\delta x^\rho = \xi^\rho(x)$ leads to the conservation of $T^{\mu \nu}$ itself, whereas the invariance under a Lorentz transformation $\delta x^\rho = \omega^\rho{}_\lambda(x) \, x^\lambda$ leads to the conservation of the {\em total} angular momentum tensor~\cite{weinberg}
\be
J^{\rho \mu \nu} = x^\mu \, T^{\rho \nu} - x^\nu \, T^{\rho \mu}.
\label{am}
\ee

Now, when the local kinematics is assumed to be ruled by the de Sitter group, the underlying spacetime is necessarily a de Sitter spacetime. As already said, that spacetime is not transitive under ordinary translations, but under the so called de Sitter ``translations'', whose infinitesimal version is
\be
\delta x^\rho = \xi^\alpha(x) \Delta_\alpha{}^\rho,
\label{dstrans}
\ee
where $\xi^\alpha(x)$ is the transformation parameter and\footnote{For ordinary translations, as is well known, the Killing vectors reduce to $\delta_\alpha^\rho$.}
\be
\Delta_\alpha{}^\rho = \delta_\alpha^\rho - \frac{1}{4l^2} \left( 2 g_{\alpha \nu} \, x^\nu x^\rho -
x^2 \delta_\alpha^\rho \right) \equiv
\delta_\alpha^\rho - \frac{1}{4l^2} \bar{\delta}_\alpha{}^\rho
\label{dsKilling}
\ee
represent the Killing vector components, with $x^2 = g_{\mu \nu} \, x^\mu x^\nu$. Under such a transformation, the metric tensor changes according to
\be
\delta g_{\mu \nu} = - \nabla_\nu [\Delta_{\alpha \mu} \xi^\alpha(x)] -
\nabla_\mu [\Delta_{\alpha \nu} \xi^\alpha(x) ],
\ee
with $\nabla_\nu$ a covariant derivative in the spacetime metric. Using the fact that the $\Delta_{\alpha \mu}$'s are Killing vectors, this can be rewritten in the form
\be
\delta g_{\mu \nu} = - \Delta_{\alpha \mu} \nabla_\nu \xi^\alpha(x) -
\Delta_{\alpha \nu} \nabla_\mu \xi^\alpha(x).
\label{metricTrans}
\ee
Substituting in Eq.~(\ref{Svar1}), the invariance of the action yields the conservation law
\be
\nabla_\mu \Pi^{\mu \nu} = 0,
\label{dSconservation}
\ee
where
\be
\Pi^{\mu \nu} \equiv T^{\mu \alpha} \Delta_\alpha{}^\nu =
T^{\mu \nu} - \frac{1}{4l^2} \, {K}^{\mu \nu},
\label{TmK}
\ee
with $T^{\mu \nu}$ the symmetric energy--momentum tensor, and $K^{\mu \nu}$ the proper conformal current~\cite{coleman}
\be
K^{\mu \nu} \equiv T^{\mu \alpha} \bar{\delta}_\alpha{}^\nu =
T^{\mu \alpha} \left(2 g_{\alpha \rho} \, x^\rho x^\nu -
x^2 \delta_\alpha{}^\nu \right).
\label{KdelT}
\ee
When the underlying spacetime is the de Sitter spacetime, the covariant conserved source is the projection of the energy--momentum tensor $T^{\mu \alpha}$ along the Killing vector $\Delta_{\alpha}{}^{\mu}$.

In general, neither $T_{\mu \nu}$ nor $K_{\mu \nu}$ is conserved separately. In fact, as an explicit calculation shows,
\be
\nabla_\mu T^{\mu \nu} = \frac{2 \, T^\rho{}_\rho \, x^\nu}{4l^2 - x^2} \qquad \mbox{and} \qquad
\nabla_\mu {K}^{\mu \nu} = \frac{2 \, T^\rho{}_\rho \, x^\nu}{1 - x^2/4l^2}.
\label{Conser1}
\ee
Only when the trace of the energy--momentum tensor vanishes are the currents $T^{\mu \nu}$ and ${K}^{\mu \nu}$  separately conserved. In the formal limit of a vanishing cosmological term (corresponding to $l \to \infty$), we obtain
\be
\nabla_\mu T^{\mu \nu} = 0 \qquad \mbox{and} \qquad
\nabla_\mu K^{\mu \nu} = 2 \, T^\rho{}_\rho \, x^\nu.
\label{Conser2}
\ee
On the other hand, in the formal limit of an infinite cosmological term (corresponding to $l \to 0$), we get
\be
\nabla_\mu T^{\mu \nu} = - \, 2 \, T^\rho{}_\rho \, \frac{x^\nu}{x^2} \qquad \mbox{and} \qquad
\nabla_\mu {K}^{\mu \nu} = 0.
\label{Conser3}
\ee
In this limit, physics becomes conformally invariant, and the proper conformal current turns out to be conserved.

\subsection{Second Bianchi Identity}

Let us consider the gravitational action functional
\be
S_g = -\, \frac{c^3}{16 \pi G} \int R_\rt \, \sqrt{-g} \; d^4x,
\label{graviaction}
\ee
with $R_\rt$ the ordinary scalar curvature generated by $T_{\mu \nu}$. Up to a surface term, the variation of this action is
\be
\delta S_g = -\, \frac{c^3}{16 \pi G} \int G_\rt^{\mu \nu} \, \delta g_{\mu \nu} \, \sqrt{-g} \, d^4x,
\label{deltaGA}
\ee
where
\be
G_\rt^{\mu \nu} = R_\rt^{\mu \nu} - \onehalf \, g^{\mu \nu} \, R_\rt
\ee
is Einstein's tensor. For the specific case of de Sitter ``translations'', in which the metric tensor transforms according to Eq.~(\ref{metricTrans}), we get
\be
\delta S_g = -\, \frac{c^3}{16 \pi G} \int \nabla_\mu \left( G_\rt^{\mu \alpha} \Delta_\alpha{}^\nu \right) \xi_\nu(x) \, \sqrt{-g} \, d^4x.
\label{Svar2}
\ee
From the invariance of the action, and considering the arbitrariness of the parameter $\xi_\nu(x)$, we obtain
\be
\nabla_\mu \left( G_\rt^{\mu \alpha} \Delta_\alpha{}^\nu \right) = 0.
\label{cBi}
\ee
In the case of ordinary general relativity, whose underlying kinematics is ruled by the Poincar\'e group, the Killing vectors are simply $\delta_\alpha^\nu$, and Eq.~(\ref{cBi}) reduces to the usual contracted form of the second Bianchi identity~\cite{cw}. When the underlying kinematics is ruled by the de Sitter group, however, what is covariantly conserved is the projection of $G_\rt^{\mu \alpha}$ along the Killing vector $\Delta_\alpha{}^\nu$. This projection represents Einstein's tensor of the {\em total} spacetime curvature: the dynamic curvature generated by $T^{\mu \nu}$, and the kinematic curvature associated to the local de Sitter spacetime. It is, for this reason, the Bianchi identity consistent with de Sitter special relativity.

\subsection{Gravitational Field Equations}

In ordinary general relativity, Einstein equation appears as an equality between two co\-vari\-antly--conserved quantities: the purely geometrical Einstein tensor --- divergenceless by the second Bianchi identity --- and the source energy-momentum tensor --- divergenceless by Noether's theorem. Consistency with de Sitter special relativity --- in which $\Pi_{\mu \nu}$ and not $T_{\mu \nu}$ is the conserved current --- requires that Einstein equation be generalized to\footnote{It should be remarked that this equation follows from the same Lagrangian of ordinary general relativity, provided the translational variation $\delta g_{\mu \nu} = \delta^\alpha_\nu \delta g_{\mu \alpha}$ be replaced by the de Sitter variation $\delta g_{\mu \nu} = \Delta^\alpha{}_\nu \delta g_{\mu \alpha}$. The name ``translational variation'' is related to the fact that this is the variation used to define the energy--momentum tensor, which is obtained from the invariance of the action under spacetime translations, with Killing vectors $\delta^\alpha_\nu$.}
\be
G_\rt^{\mu \alpha} \Delta_\alpha{}^\nu = \frac{8 \pi G}{c^4} \, T^{\mu \alpha} \Delta_\alpha{}^\nu.
\ee
Substituting $\Delta_\alpha{}^\nu$ from Eq.~(\ref{dsKilling}) and using Eq.~(\ref{TmK}), it becomes
\be
G_\rt^{\mu \nu} - \frac{1}{4l^2} \, G_\rt^{\mu \alpha} \bar{\delta}_\alpha{}^\nu =
\frac{8 \pi G}{c^4} \left( T^{\mu \nu} - \frac{1}{4l^2} \, K^{\mu \nu} \right).
\label{nee}
\ee
This field equation has both a symmetric and an anti--symmetric parts. Using the relations
\be
K_\a^{\mu \nu} \equiv \onehalf K^{[\mu \nu]} = \onehalf \left(K^{\mu \nu} - K^{\nu \mu} \right) =
x^\rho \left(x^\mu T_\rho{}^\nu - x^\nu T_\rho{}^\mu \right)
\ee
and
\be
K_\s^{\mu \nu} \equiv \onehalf K^{(\mu \nu)} =  \onehalf \left(K^{\mu \nu} + K^{\nu \mu} \right) =
x^\rho \left(x^\mu T_\rho{}^\nu + x^\nu T_\rho{}^\mu -x_\rho T^{\mu \nu} \right),
\ee
the anti--symmetric part brings back the total angular momentum tensor,
\be
x^\rho \left(x^\mu G_{\rt \rho}{}^\nu - x^\nu G_{\rt \rho}{}^\mu \right) = \frac{8 \pi G}{c^4}
x^\rho J_\rho{}^{\mu \nu},
\ee
from where we see that de Sitter relativity does not produce any change in the rotational --- or Lorentzian --- sector of the theory. The symmetric part, on the other hand, yields
\be
G_\rt^{\mu \nu} - G_\rk^{\mu \nu} =
\frac{8 \pi G}{c^4} \left( T^{\mu \nu} - \frac{1}{4l^2} \, K_\s^{\mu \nu} \right),
\label{nee2}
\ee
where, analogously to (\ref{KdelT}), we have identified
\be
\frac{1}{l^2} \, G_\rt^{(\mu \alpha} \bar{\delta}_\alpha{}^{\nu)} \equiv G_\rk^{\mu \nu} = 
R_\rk^{\mu \nu} - \onehalf \, g^{\mu \nu} \, R_\rk,
\ee
with $G_\rk^{\mu \nu}$ representing the kinematic curvature related to the local ``de Sitter--like'' spacetime, necessary to yield the appropriate local symmetry, as required by the de Sitter version of the strong equivalence principle.

By ``de Sitter--like'' we mean a spacetime whose curvature tensor is formally the same as the curvature tensor of a de Sitter spacetime, except for the facts that (i) $\Lambda$ is no longer constant, and (ii) it is written with the general spacetime metric $g_{\mu \nu}$, solution of the complete field equation (\ref{nee2}). Namely,
\be
R_\rk^{\mu}{}_{\nu \rho \sigma} = - \, \frac{\Lambda}{3} \,
\left(\delta^{\mu}_{\rho} \, g_{\nu \sigma} - \delta^{\mu}_{\sigma} \, g_{\nu \rho} \right).
\label{47}
\ee
The corresponding Ricci tensor and the scalar curvature have, consequently, the forms
\be
R_\rk^{\mu \nu} = - \, \Lambda \, g^{\mu \nu} \quad \mbox{and} \quad
R_\rk = - \, 4\, \Lambda.
\label{rlambda}
\ee
If the dynamic gravitational field related to general relativity is neglected, it reduces to a pure de Sitter spacetime. Using the above expressions, the generalized Einstein equation (\ref{nee2}) can be rewritten as
\be
R_\rt^{\mu \nu} - \onehalf \, g^{\mu \nu} \, R_\rt - \frac{\Lambda}{4} \, g^{\mu \nu} =
\frac{8 \pi G}{c^4} \left( T^{\mu \nu} - \frac{1}{4l^2} \, K_\s^{\mu \nu} \right).
\label{nee3}
\ee
By construction, this equation is consistent with {\it de Sitter special relativity}, and for this reason the corresponding gravitational theory can be called {\it de Sitter general relativity}. In the limit $\Lambda \to 0$ (which corresponds to $l \to \infty$), the field equation (\ref{nee3}) reduces to the usual Einstein equation (\ref{ufee}), which is consistent with ordinary (Poincar\'e) special relativity. Notice that $\Lambda$ is no longer a constant. This is clear from the fact that it is non--vanishing in the region occupied by the source, and goes to zero outside that region.

\subsection{Conformal Current and the Cosmological Term}
\label{sec:res}

Neglecting both the source and the corresponding gravitational field related to ordinary general relativity --- as usually done in special relativity --- the (trace of the) field equation (\ref{nee3}) yields 
\be
\Lambda = \frac{2 \pi G}{c^4} \, \frac{K^\mu{}_{\mu}}{l^2}.
\label{***}
\ee
According to de Sitter relativity, therefore, the source of the cosmological term is the trace 
\be
K^\mu{}_{\mu} =  2 \, x^\mu x^\nu T_{\mu \nu} - x^2 \, T^{\mu}{}_{\mu}
\label{traceofK}
\ee
of the proper conformal current of matter. It is important to note that, differently from the usual de Sitter solution, the cosmological term here is produced by ordinary matter. In absence of matter, the proper conformal current vanishes, and the cosmological term vanishes as well.

Let us consider now the specific case of an isotropic and homogeneous dust fluid described by
\be
T_{\mu \nu} = \varepsilon \, u_\mu u_\nu.
\ee
Considering that the underlying de Sitter spacetime is described by the metric
\be
g_{\mu \nu} = \Omega^2 \, \eta_{\mu \nu},
\label{udsmetric}
\ee
with $\Omega$ given by Eq.~(\ref{n}), it is easy to see that, in a comoving coordinate system,
\be
K^\mu{}_\mu = \left(c^2 t^2 + r^2 \right) \varepsilon,
\ee
where $r^2 = \delta_{ij} x^i x^j$. Interesting enough, the trace of the proper conformal current produces a Euclidianization of the coordinate--dependent factor, yielding a strictly positive quantity. At the Planck scale, $t = t_P$, $r = l_P$ and $\varepsilon = \varepsilon_P$, which yields
\be
K^\mu{}_{\mu} = 2 \, l^2_P \, \varepsilon_P.
\ee
At an arbitrary energy scale, therefore, the trace of the proper conformal current is 
\be
K^\mu{}_{\mu} = 2 \, l^2 \varepsilon.
\label{camimi}
\ee
Substituting in Eq.~(\ref{***}), we get
\be
\Lambda = \frac{4 \pi G}{c^4} \, \varepsilon,
\label{kineLambdaBis}
\ee
which is again Eq.~(\ref{kineLambda}), now obtained from its relation with the proper conformal current of the source.

\section{The Cosmological Constant Problem}

Although supposed to be relevant at ultra--high energy--densities, de Sitter relativity may give rise to residual effects at not so high energies. As an example, let us consider the universe as a whole. Of course, since the gravitational field equations of de Sitter general relativity are slightly different from the usual ones, the ensuing Friedmann equations will also be modified. However, for the sake of simplicity, we assume that spacetime is still described by the Friedmann-Robertson-Walker metric, and has a flat space--section ($k=0$). In this case, we can take $\varepsilon$ of the order of the Friedmann critical energy density
\be
\varepsilon \simeq \frac{3 H_0^2 c^2}{8 \pi G},
\ee
with $H_0$ the Hubble constant. Substituting in Eq.~(\ref{kineLambdaBis}), we obtain
\be
\Lambda \simeq \frac{3 H_0^2}{2 c^2}.
\label{coscon2}
\ee
Using the value $H_0 = 75$~(Km/s)/Mpc, the cosmological constant is found to be
\be
\Lambda \simeq 10^{-56}~\mbox{cm}^{-2},
\ee
which is of the order of magnitude of the observed value~\cite{obs}. This simple estimate is given to illustrate the main point:  replacing Poincar\'e by de Sitter as the group governing the spacetime kinematics leads to a relation between the energy density of any physical system and the local value of $\Lambda$. When applied to the whole universe, that relation is able to predict the value of the cosmological constant,  which is no more an independent parameter --- it is determined by the spacetime kinematics and is, in principle, calculable. It is important to observe that, although produced by matter, the source of $\Lambda$ is not the energy--momentum tensor, but the proper conformal current. In consequence, no mysterious fluid --- satisfying exotic equations of state --- is necessary to explain the existence of a cosmological constant.

According to de Sitter relativity, therefore, the cosmological term has a pure kinematic origin. In this case, the thermodynamic energy density associated with the de Sitter horizon is found to be \cite{abap}
\be
\varepsilon_\Lambda =
\frac{c^4}{4 \pi G} \, \Lambda.
\ee
A comparison with Eq.~(\ref{***}) yields
\be
\frac{K^\mu{}_\mu}{l^2} = 2 \, \varepsilon_\Lambda.
\label{61}
\ee
Assuming for argument's sake that the present-day content of the universe can be accurately described by dust, we see from Eqs.~(\ref{camimi}) and (\ref{61}) that the dark and the matter energy densities of such universe are of the same order, $\varepsilon_\Lambda \simeq \varepsilon$, a result consistent with the so-called cosmic coincidence problem~\cite{carroll}.
  
\section{Photon Kinematics in de Sitter Relativity}

According to quantum gravity considerations, high energies might cause small--scale fluctuations in the texture of spacetime. These fluctuations could, for example, act as small--scale lenses, interfering in the propagation of ultra--high energy photons. The higher the photon energy, the more it changes the spacetime structure, the larger the interference will be. This kind of mechanism could be the cause of the recently observed delay in high energy gamma--ray flares from the heart of the galaxy Markarian 501~\cite{magic}. Those observations compared gamma rays in two energy ranges, from 1.2 to 10~TeV, and from 0.25 to 0.6~TeV. The first group arrived on Earth four minutes later than the second. Since de Sitter relativity gives a precise meaning to these local spacetime fluctuations, it provides a precise high energy phenomenology, opening up the door for experimental predictions.

With this in mind, let us consider a photon of wavelength $\lambda$ and energy $E = h c/ \lambda$. Although the photons in a gamma--ray beam are not necessarily in thermal equilibrium, we are going to use the thermodynamic expression~\cite{stat}
\be
\varepsilon = \frac{\pi^2}{15} \, \frac{(k T)^4}{(\hbar c)^3}
\ee
to estimate the photons energy density. Setting $k T = E$, it becomes
\be
\varepsilon = \frac{\pi^2}{15} \, \frac{E^4}{(\hbar c)^3}.
\label{phenden}
\ee
Substituting in Eq.~(\ref{kineLambdaBis}), we obtain
\be
\Lambda \simeq \frac{4 \pi^3}{15 \hbar^2 c^2} \, \frac{E^4}{E^2_P},
\label{gammaphoton}
\ee
where $E_P = \sqrt{c^5 \hbar/G}$ is the Planck energy. The corresponding de Sitter length parameter is given by
\be
l = \sqrt{3/\Lambda}.
\label{dSlp}
\ee

To get an idea of the order of magnitude, we give in Table 1 the local values of $l$ and $\Lambda$ for several photons with different wavelength $\lambda$. In the first line are the values for a photon with energy of the order of the Planck energy. Gamma-rays (1) and (2) correspond to the two  observed gamma-ray flares  from  Markarian 501. For comparison purposes, we give also the values for a visible (red) photon.
\begin{table}[ht]
\begin{center}
\begin{tabular}{|l|l|l|l|l|}
\hline
 & ~~$E$ (GeV) & ~~~${\lambda}$ (cm) & ~~~${l}$ (cm) & ~$\Lambda$ (cm$^{-2}$) \\ 
\hline \hline
Planck photon & $1.2 \times 10^{19}$ & $1.0 \times 10^{-32}$ & $9.7 \times 10^{-34}$ &
$3.3 \times 10^{66}$\\
\hline
Gamma-ray (1) & $1.0 \times 10^{4}$ & $1.2 \times 10^{-17}$ & $1.4 \times 10^{-3}$ &
$1.7 \times 10^{6}$\\
\hline
Gamma-ray (2) & $0.6 \times 10^{3}$ & $2.1 \times 10^{-16}$ & $3.8 \times 10^{-1}$ &
$2.2 \times 10^{1}$\\
\hline
Red light & $ 1.8 \times 10^{-9}$ & $7.0 \times 10^{-5}$ & $4.5 \times 10^{22}$ &
$1.6 \times 10^{-45}$\\
\hline
\end{tabular}
\caption{\it Local values of $l$ and $\Lambda$ for several different photons.}
\end{center}
\end{table}
Since the photons produce such $\Lambda$ in the place they are located, we can assume that they are always propagating in a de Sitter spacetime with that cosmological term.

\subsection{Geometric Optics Revisited}

In flat spacetime, the condition for geometric optics to be applicable is that
\be
\lambda \ll l,
\label{goc}
\ee
where $l$ is the typical dimension of the physical system. Since the physical system is now the local de Sitter spacetime produced by the photon, that dimension is given by the de Sitter length parameter $l$. From Table 1 we see that, for a photon with wavelength of the order of the Planck length, this condition is not fulfilled. However, for gamma-rays (1) and (2), as well as for red light, condition (\ref{goc}) is fulfilled, which means that we can use geometric optics to study their propagation.

In the geometric optics domain, any wave-optics quantity $A$ which describes the
wave field is given by an expression of the type
\be
A = b \, {\rm e}^{i \phi},
\ee
where the amplitude $b$ is a slowly varying function of the coordinates and time, and the
phase $\phi$, the eikonal, is a large quantity which is {\it almost linear} in the
coordinates and the time. The time derivative of $\phi$ yields the angular frequency of the
wave,
\be
\frac{\partial \phi}{\partial t} = \omega,
\ee
whereas the space derivative gives the wave vector
\be
\frac{\partial \phi}{\partial \mbox{\boldmath$r$}} = - \, \mbox{\boldmath$k$}.
\ee
The characteristic equation for Maxwell's equations in an isotropic 
(but not necessarily homogeneous) medium of refractive index $n(r)$ is
\be
\left(\frac{\partial \phi}{\partial \mbox{\boldmath$r$}}\right)^2 -
\frac{n^2(r)}{c^2} \, \left(\frac{\partial \phi}{\partial t}\right)^2 = 0 \; ,
\label{eiko1}
\ee
which implies the usual relation
\be
\mbox{\boldmath$k$}^2 = n^2(r) \, \frac{\omega^2}{c^2}.
\ee

Now, as is well known, there exists a deep relationship between optical media and metrics~\cite{Mol52}. This relationship allows to reduce the problem of the propagation of electromagnetic waves in a gravitational field to the problem of wave propagation in a refractive medium in flat spacetime. Let us then consider the specific case of a de Sitter spacetime, for which the quadratic line element $ds^2$ can be written in the form~\cite{ellis}
\be
ds^2 = d \tau^2 - n^2(E) \, \delta_{i j} \, dx^i dx^j,
\label{dss}
\ee
where
\be
n(E) \equiv \exp\left[\sqrt{\Lambda/3} \; \tau  \right],
\label{idr0}
\ee
with $\tau = c \, t$. In these coordinates, the metric components are
\be
g_{0 0} = g^{0 0} = 1, \quad
g_{i j} = - \, n^2(E) \, \delta_{i j},
\label{rime}
\ee
and the components of the ``conformal'' Ricci tensor is
\be
R_{\rk}^\mu{}_\nu = - \, {\Lambda} \, \delta^\mu{}_\nu,
\label{ricci2}
\ee
with $\Lambda$ given by Eq.~(\ref{gammaphoton}). It is then easy to see that, with the metric components~(\ref{rime}), the curved spacetime eikonal equation for a $n=1$ refractive medium,
\be
g^{\mu \nu} \, \frac{\partial \phi}{\partial x^\mu} \,
\frac{\partial \phi}{\partial x^\nu} = 0,
\label{eiko2}
\ee
coincides formally with the flat-spacetime eikonal equation (\ref{eiko1}), valid in a
medium of refractive index $n(r)$. For this reason, $g_{i j}$ is usually called the
{\it refractive metric}, with $n(E)$ playing the role of refractive index~\cite{livro}.

\subsection{Electromagnetic Waves in the Geometric Optics Limit}

As already remarked, according to de Sitter relativity the photons produce a local de Sitter spacetime in the place they are located. We can then assume that they are always propagating in a de Sitter spacetime, with $\Lambda$ given by Eq.~(\ref{gammaphoton}). Let us then consider the electromagnetic field equations in a de Sitter spacetime, restricting ourselves to the domain of geometric optics. Denoting the electromagnetic gauge potential by $A_\mu$, and assuming the generalized Lorenz gauge $\nabla_\mu A^\mu=0$, Maxwell's equation reads
\be
\Box A^\mu - R_{\rk}^{\mu}{}_{\nu} \, A^\nu = 0,
\label{me2}
\ee
where $\Box=g^{\lambda \rho} \nabla_\lambda \nabla_\rho$. Substituting the Ricci tensor components (\ref{ricci2}), Maxwell equation (\ref{me2}) becomes
\be
\Box A^\mu + {\Lambda} \, A^\mu = 0.
\label{me3}
\ee
Although the term involving the cosmological constant looks like a background-dependent mass for the photon field, this interpretation leads to properties which are physically unacceptable~\cite{faraoni}. In fact, as Maxwell equations in four dimensions are conformally invariant, and de Sitter spacetime is conformally flat, the electromagnetic field must propagate on the light--cone~\cite{lico}, which implies a vanishing mass for the photon field.

Assuming a massless photon field, therefore, we take the monochromatic plane-wave
solution to the field equation (\ref{me3}) to be
\be
A_\mu = b_\mu \exp[i \, k_\alpha \, x^\alpha],
\label{mws}
\ee
where $b_\mu$ is a polarization vector, and $k_\alpha = ({\omega(|\mbox{\boldmath$k$}|)}/{c}, - \, \mbox{\boldmath$k$})$ is the wave-number four-vector, with $\omega(|\mbox{\boldmath$k$}|)$ the angular frequency. In order to be a solution of equation (\ref{me3}), the following dispersion relation must be satisfied,
\be
\omega(k) = \frac{c}{n(E)} \, \left[ k^2 +
{n^2(E) \, \Lambda} \right]^{1/2}, 
\label{redi}
\ee
where we used the notation $k = |\mbox{\boldmath$k$}|$. Considering that
\be
\frac{1}{n(E) \, \Lambda^{1/2}} \sim l,
\ee
with $l$ the dimension of the local de Sitter spacetime, and remembering that $k \sim
\lambda^{-1}$, the condition (\ref{goc}) for geometric optics to be applicable
turns out to be
\be
k \gg n(E) \, \Lambda^{1/2}.
\ee
In this domain, therefore, the dispersion relation (\ref{redi}) assumes the form
\be
\omega(k) = c \, \frac{k}{n(E)}, 
\label{redis}
\ee
and the corresponding velocity of propagation of an electromagnetic wave, given by the
group velocity, is~\cite{walter}
\be
v \equiv \frac{d \omega(k)}{d k} = \frac{c}{n(E)}.
\label{gv}
\ee
In the limit $\Lambda \to 0$, which corresponds to a contraction from de Sitter to ordinary special relativity, $n(E) \to 1$, and there will be no effect on the photon propagation.

\subsection{Application to the Gamma--Ray Flares}

Let us consider now the propagation of gamma--rays. Substituting the local cosmological term~(\ref{gammaphoton}) into the refractive index (\ref{idr0}), we obtain
\be
n(E) \simeq \exp \left[\sqrt{\frac{4 \pi^3}{45 \hbar^2 c^2}} \; \frac{E^2}{E_P} \; \tau \right].
\label{idr1}
\ee
For the local de Sitter spacetime produced by a photon, the length $\tau$ can be identified with its own wavelength $\lambda = h c/E$. Hence, we get
\be
n{(E)} \simeq \exp \left[\sqrt{\frac{16 \pi^5}{45}} \; \frac{E}{E_P} \right].
\ee
For energies small compared to $E_P$, we can write
\be
n{(E)} \simeq 1 + \sqrt{\frac{16 \pi^5}{45}} \, \frac{E}{E_P}.
\ee
For a visible (red) electromagnetic radiation,
\be
n_{(\mbox{\footnotesize{red}})} \simeq 1 + 1.9 \times 10^{-27}.
\ee
For gamma--rays (1) and (2), we get, respectively,
\be
n_{(1)} \simeq 1 + 8.8 \times 10^{-15} \quad \mbox{and} \quad
n_{(2)} \simeq 1 + 5.2 \times 10^{-16}.
\label{refractionindex}
\ee
Taking into account that the velocity of each photon is given by Eq.~(\ref{gv}), the time difference $\Delta t$ to travel a distance $d$ will be
\be
\Delta t = \frac{d}{c}  \left[n_{(1)} - n_{(2)} \right].
\ee
Using the refractive indices (\ref{refractionindex}), we see that, for a distance of 500 millions light--year, which corresponds to $d = 4.7 \times 10^{26}$~cm, the time difference will be
\be
\Delta t \simeq 130~\mbox{s} = 2.2~\mbox{min}.
\ee
This is  of the same order of magnitude of the observed delay between the two gamma--ray flares originated from the center of the galaxy Markarian 501~\cite{magic}.

\section{Final Remarks}

There are theoretical and experimental evidences that ordinary special relativity, whose underlying kinematics is ruled by the Poincar\'e group, breaks down at ultra--high energy densities. When looking for a new special relativity, the most natural generalization is arguably to replace Poincar\'e special relativity by a de Sitter special relativity. This means to assume that the local kinematics is ruled by the de Sitter group. This, in turn, means that any physical process must modify the local structure of spacetime in such a way that the region where the process takes place departs from Minkowski and becomes an asymptotically flat de Sitter spacetime. 

Considering that the de Sitter spacetime is transitive under a combination of translations and proper conformal transformations, the proper conformal transformations will naturally be incorporated in the spacetime kinematics. As an immediate consequence, the source in the gravitational field equation will be a combination of energy--momentum and proper conformal currents: whereas the energy--momentum tensor appears as source of ordinary curvature, the proper conformal current $K_{\mu \nu}$ appears as source of the local de Sitter spacetime, which is necessary to comply with the local kinematics, now governed by the de Sitter group. In this theory, therefore, the existence of $\Lambda$ is directly connected with the conformal sector of the spacetime kinematics.

When applied to the whole universe, de Sitter special relativity is able to predict, from the current matter content of the universe, the value of $\Lambda$. It gives, furthermore, an explanation for the cosmic coincidence problem. When applied to study the propagation of ultra--high energy photons, it gives a good estimate for the recently observed delay in high energy gamma--ray flares coming from the center of the galaxy Markarian 501~\cite{magic}. If this delay is a manifestation of the small--scale fluctuations in the texture of spacetime, predicted to exist at very high energies, de Sitter relativity can be seen as a new paradigm to approach quantum gravity. Of course, the experimental evidences of the above delays are still very fragile. Independently of this fact, de Sitter special relativity predicts the existence of such delay.

Even though conformal symmetry is not an exact symmetry at low energies, according to de Sitter special relativity it naturally becomes the relevant symmetry at ultra--high energy densities. In fact, the higher the energy density, the higher the value of $\Lambda$, the higher the importance of conformal symmetry. Near the Planck energy, the local value of $\Lambda$ will be very large, and the local de Sitter space will approach a cone--spacetime, which is transitive under proper conformal transformations only~\cite{dssr}. Under such extreme conditions, physics becomes conformal invariant, and the proper conformal current will be conserved (see the conservation laws (\ref{Conser1}), as well as their high-energy limit (\ref{Conser3})). An interesting property of this geometrical structure is that the cone--spacetime is a kind of dual to Minkowski, with the duality transformation given by the spacetime inversion~\cite{abap}
\be
x^a \to - \, \frac{x^a}{\sigma^2} \,.
\label{sti}
\ee
The same happens to the corresponding transitivity generators: under the spacetime inversion (\ref{sti}), the proper conformal generators --- which define the transitivity on the cone spacetime --- are transformed into  the translation generators~\cite{coleman} --- which define the transitivity on Minkowski spacetime. This duality symmetry between high and low energies may have important consequences for high--energy physics and, in particular, for quantum gravity.

\section*{Acknowledgments}
The authors would like to thank J. P. Beltr\'an Almeida and C. S. O. Mayor for useful discussions. They would like to thank also FAPESP, CNPq and CAPES for partial financial support.


\end{document}